\documentclass[twoside,a4paper,11pt]{sea10}
\usepackage{graphicx}
\usepackage{hyperref}
\begin{document}
\pagenumbering{arabic}
\pagestyle{myheadings}
\thispagestyle{empty}
{\flushleft\includegraphics[width=\textwidth,bb=58 650 590 680]{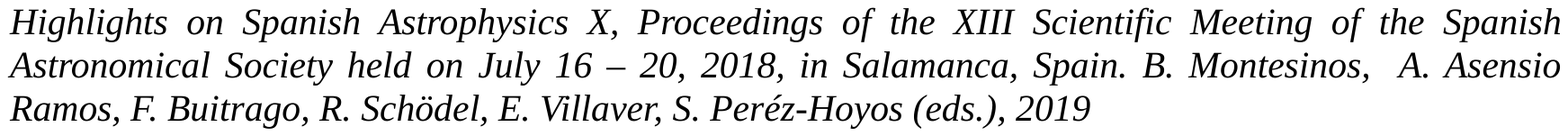}}
\vspace*{0.2cm}
\begin{flushleft}
{\bf {\LARGE
%
The Stellar Tidal Stream Survey 
%
}\\
\vspace*{1cm}
%
David Mart\'\i nez-Delgado$^{1}$
%
}\\
\vspace*{0.5cm}
%

$^{1}$Astronomisches Rechen-Institut, Zentrum f\"ur Astronomie der
Universit\"at Heidelberg, Heidelberg, Germany\\

%
\end{flushleft}
%
\markboth{
Tidal Stream Survey
}{ 
%
Mart\'\i nez-Delgado
%
}
\thispagestyle{empty}
\vspace*{0.4cm}
\begin{minipage}[l]{0.09\textwidth}
\ 
\end{minipage}
\begin{minipage}[r]{0.9\textwidth}
\vspace{1cm}
\section*{Abstract}{\small
%

Mergers and tidal interactions between massive galaxies and their
dwarf satellites are a fundamental prediction of the Lambda-Cold Dark
Matter cosmology. These events are thought to influence galaxy
evolution throughout cosmic history and to provide important
observational diagnostics of structure formation. Stellar streams in the Local Group are spectacular evidence for satellite disruption at the present day. However, constructing a significant sample of tidal
streams beyond our immediate cosmic neighborhood has proven a daunting
observational challenge and their potential for deepening our
understanding of galaxy formation has yet to be realized. Over the last decade, the Stellar Tidal Stream Survey has obtained deep, wide-field images of nearby Milky-Way analog galaxies with a network of robotic amateur telescopes, revealing for the first time an assortment of large-scale tidal structures in their halos. I discuss the main results of this project and future plans for performing dynamical studies of the discovered streams.
\normalsize}
\end{minipage}
%
%
%
\section{Stellar Tidal Streams as Galaxy Formation Diagnostic}

Within the hierarchical framework for galaxy formation, the stellar bodies of massive galaxies are expected to form and evolve not only through the inflow of cold gas, but also the infall and successive mergers of low-mass, initially bound systems. Commonly referred to as satellites, they span a wide mass range and consist of dark matter, gas, and, in most cases, stars.  While the interaction rate is expected to drop to the present-day epoch, numerical cosmological models, built within the  $\Lambda$-Cold Dark Matter (LCDM) paradigm (e.g. Bullock \& Johnston 2005; Cooper et al. 2010), predict that such satellite disruption still occurs around all massive galaxies. As a consequence, the stellar halos of these galaxies should contain a wide variety of diffuse structural features, such as {\bf stellar streams} or shells, that result from interactions and mergers with dwarf satellites. The most spectacular cases of tidal debris  are long, dynamically cold stellar streams,  formed from a disrupted dwarf satellite, that wrap around the host galaxy's disk and roughly trace the orbit of the progenitor satellite. Although these fossil records disperse into amorphous clouds of debris (through phase-mixing) in a few Giga-years,  LCDM simulations predict that stellar streams may be detected nowadays, with sufficiently deep observations, in the outskirts of almost all nearby galaxies.

\begin{figure}[t]
\center
\includegraphics[scale=0.45]{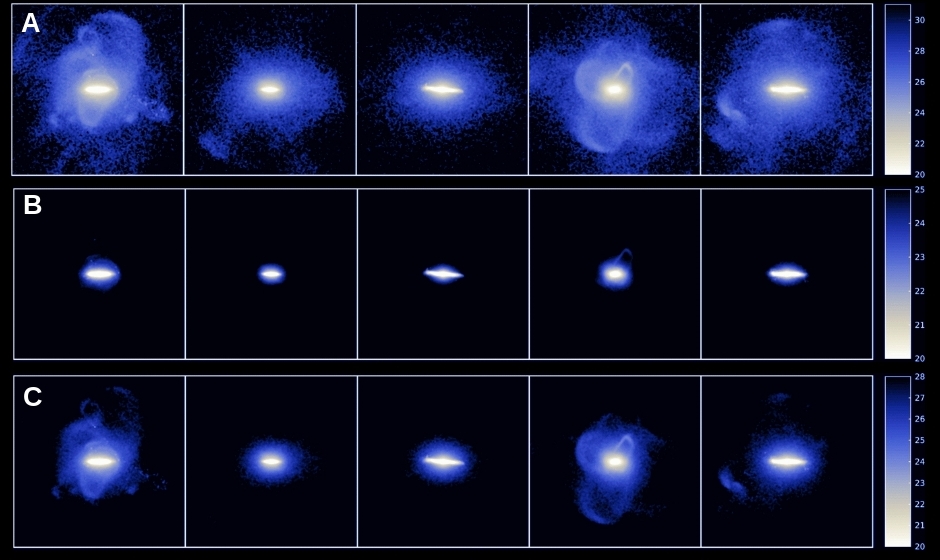}
 \caption{\looseness=-1 Expected `halo streams' around a Milky Way - like
galaxy from the Auriga cosmological simulations (Grand et al. 2017). The panels
show an external perspective of several realizations of a simulated galaxy within the hierarchical framework, with streams resulting from tidally 
disrupted satellites. They illustrate a variety of typical accretion histories for Milky Way-type galaxies. Each panel is 300~kpc on a side. The different rows show theoretical  predictions for detectable tidal features in each halo model, assuming three different surface brightness (SB) limit detection limits ({\it A}: $\mu_{\mathrm{lim}}=31$, {\it B}:$\mu_{\mathrm{lim}} = 25$ and {\it C}: $\mu_{\mathrm{lim}}= 28$
mag/arcsec$^2$). 
This suggests that the number of tidal features visible in the outskirts of spirals varies dramatically with the SB limit of the data, with no discernible
sub-structure expected for surveys with SB limits brighter than $\sim25$ mag/arcsec$^2$ (e.g.\ POSS-II and SDSS).}
\end{figure}

\vspace{0.2cm}
The detection of these faint tidal remnants is a ubiquitous aspect of galaxy formation that has not yet been fully exploited, mainly because they are challenging to observe. Although the most luminous examples of diffuse stellar streams and shells  around massive elliptical galaxies have been known for many decades, recent studies have showed that fainter analogues of these structures are common around spiral galaxies in the local universe, including the Milky Way and Andromeda (Belokurov et al. 2006; Ibata et al. 2007). 
These observations provide sound empirical support for the $\Lambda$CDM prediction that tidally disrupted dwarf galaxies could be important contributors to the stellar halo formation in the Local Group spirals.  

\begin{figure}[ht!]
\center
\includegraphics[scale=0.5]{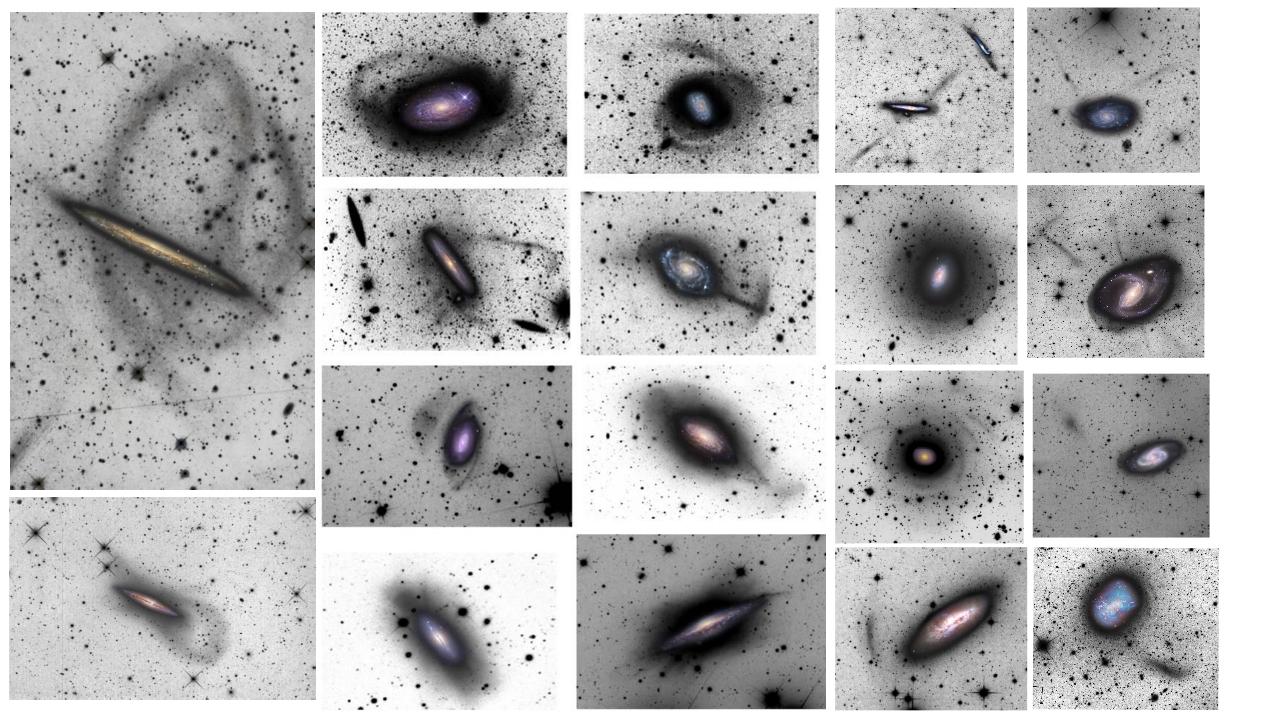}
\caption{\looseness=-1 Luminance filter images of nearby galaxies from the Stellar Tidal Stream Survey showing large, diffuse light structures in their outskirts. A color inset of the disk of each galaxy has been overplotted for reference. An illustrative comparison of some of these features to the surviving structures visible in cosmological simulations is given in Martinez-Delgado et al.\ (2010; their Fig.\ 2). The typical surface brightness (r-band)  of these streams is as faint as 26 mag/arcsec$^2$. All these images were taken with robotic amateur telescopes with an aperture range of 0.1--0.5-meter.}
\end{figure}

While stellar streams in the Milky Way and Andromeda can be studied in detail, comparison with cosmological models is limited by `cosmic variance', the differences in the individual dynamical pre-histories of overall similar galaxies.  A search for analogues to these galactic fossils in a larger sample of nearby galaxies is required to understand if the recent  merging histories of the  Local Group spirals are `typical', an issue that remains unclear.  
However, in contrast to detailed predictions from simulations, the observational portrait of minor accretion events is far from complete, owing primarily to the inherent difficulty of detecting low surface brightness features. The current LCMD numerical simulations can guide this quest for star-stream observational signatures (e.g. Johnston et al. 2008; Cooper et al. 2010). Recent simulations  have demonstrated that the characteristics of sub-structure currently visible in the stellar halos are sensitive to recent (0-8 Gyr ago) merger histories of galaxies, over a timescale that corresponds to between the last few tens of percent of mass accretion for a spiral galaxy like the Milky Way. These models predict that a survey of $\sim$100 parent galaxies reaching a surface brightness of  $\sim$29 mag/arcsec$^2$ would reveal many tens of tidal features, perhaps nearly one  detectable stream per galaxy.  However, a direct comparison of these simulations with actual observations is not yet possible because no suitably deep data sets exist.

\begin{figure}[t]
\center
\includegraphics[scale=0.40]{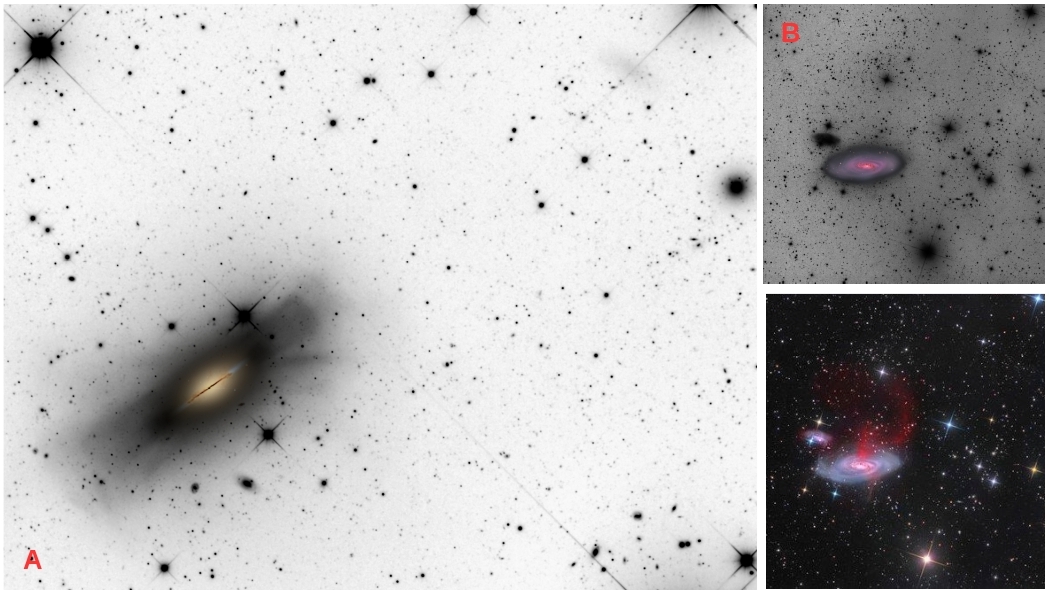}
\caption{\looseness=-1 ({\it Panel A}): The complex tidal structures of the halo of NGC 5866 (together with a tidal disrupted satellite), detected for the first time in our pilot survey in 2010, revisited with a modern commercial CCD camera in 2018. Image taken by Adam Block with the Mount Lemmon 0.8-m telescope. ({\it Panel B}): An example of very faint diffuse light detected around NGC 4569 (top panel) associated to the presence of ionized gas emission in its halo, as revealed in our follow-up deep H$_{\alpha}$ imaging (red color in the image displayed in the bottom panel). Both images were taken with a 0.5-m telescope by Mark Hanson.}
\end{figure}

\section{The Stellar Stream Survey: the first decade}

Stellar tidal streams around nearby galaxies cannot be resolved into stars with our modest telescopes and thus appear as elongated diffuse light regions that extend over several arc minutes as projected  on the sky. Our survey has established a search strategy that can successfully map such tidal streams to extremely faint surface brightness limits. Their typical surface brightness is 26 mag/arcsec$^{2}$ or fainter, depending on the luminosity of the progenitor and the time they were accreted (Johnston et al. 2001).  Detecting these faint features requires very dark sky conditions and wide-field, deep images taken with exquisite flat-field quality over a wide region ($>$ 30 arcmin) around the targets. 

The observations of the Stellar Tidal Stream Survey (STSS) are conducted with ten privately owned observatories equipped with modest-sized telescopes (0.1-0.8-meter) equipped with a latest generation astronomical commercial CCD camera and located in Europe, the United States and Chile. Each observing location features spectacularly dark, clear skies with seeing below 1.5''. The survey strategy strives for multiple deep exposures of each target using high throughput clear filters with near-IR cut-off, known as luminance (L) filters (4000 \AA $< \lambda <$7000 \AA) and a typical exposure times of  7-8 hours. Our typical 3-$\sigma$ SB detection limit (measured in random 2" diameter apertures)  is $\sim$ 28 and 27.5 mag/arcsec$^2$ in $g$ and $r$ respectively, which is approximately two magnitudes deeper than the Sloan Digital Sky Survey (SDSS) DR8 images.

We have devised a set of straightforward sample selection criteria similar to that used in the SAGA survey (Geha et al. 2017):  isolated Milky Way analog galaxies within 40 Mpc that have an $K$-band absolute magnitude M$_{K}<$ -19.6  and lying more than 20 degrees above the Galactic plane (avoiding cirrus dust and high stellar density fields). The luminosity range was selected to ensure that our sample included a significant number of Milky-Way analogue systems. Several targets that satisfied this criteria were not included, such as galaxies lying less than 20 degrees from the Galactic plane or galaxies with clear signatures of current major mergers.

 Our observational effort has also revealed previously undetected stellar streams thus making it the largest sample of tidal structures outside the Local Group. The most conspicuous examples are displayed in Figure~2. Our collection of galaxies presents an assortment of tidal phenomena exhibiting striking morphological characteristics consistent with those predicted by cosmological models. For example, in addition to identifying {\it great-circles} features that resemble the Sagittarius stream surrounding our Galaxy (e.g NGC 5907, Mart\'\i nez-Delgado et al. 2008), our observations uncovered enormous structures resembling an open umbrella that extends tens of kilo-parsecs into the halos of the spiral. These structures are often located on both sides of the host galaxy, and display long narrow shafts that terminate in a giant shell of debris (e.g. NGC 4651; Foster et al. 2014). We have also found isolated shells, giant clouds of debris floating within galactic halos, jet-like features emerging from galactic disks  and large-scale diffuse structures that are possibly related to the remnants of ancient, already thoroughly disrupted satellites. Together with these remains of possibly long-defunct companions, our data also captured surviving satellites caught in the act of tidal disruption, displaying long tails departing from the progenitor satellite. We also detected a stellar tidal stream in the halo of NGC 4449, an isolated dwarf irregular galaxy analogue to the Large Magellanic Cloud (Mart\' \i nez-Delgado et al. 2012). This appears to be the lowest-mass primary galaxy with a verified stellar stream so far. 
 This discovery suggests that satellite accretion can play a significant role in building up the stellar halos of low-mass galaxies, and possibly triggering their starburst.
 
 Our current sample so far comprises $\sim$ 50 confirmed stellar streams with a large variety of morphological types.\footnote{Only three external tidal streams  were known when the pilot survey started in 2007. We also found a large number of negative detections, in agreement with the estimated low frequency of streams at this low surface brightness regime (Morales et al. 2018).} The promising results of this foray advocate a comprehensive study of tidal streams in the nearby universe and a more insistent attention to this brand new way of understanding galaxy formation.



\begin{figure}[ht!]
\center
\includegraphics[scale=0.50]{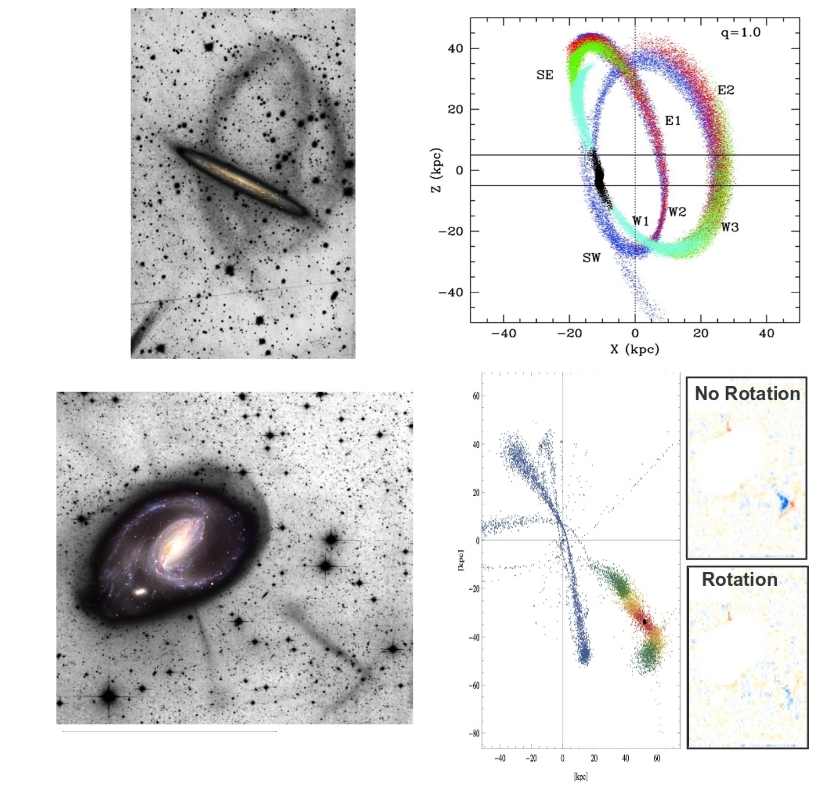} 
\caption{\looseness=-1 {({\it Top panel}): N-body simulations of the multiple wraps of the NGC 5907 tidal stream (Mart\'\i nez-Delgado et al. 2008).  Different colors correspond to the escape time of the particles. All the visible structure is explained by the destruction of a single satellite accreted $\sim$ 3.5 Gyr ago.   ({\it Bottom panel}): A spray-particle model fitting to our luminance-filter image of the NGC 1097 tidal stream (Amorisco et al.\ 2015). The stellar material shedded by the progenitor is modelled using a purposely tailored modification of the {\it particle-spray} method (Gibbons et al.\ 2014). This technique faithfully reproduces the debris of an $N$-body disruption by ejecting particles from the Lagrange points of the progenitor, allowing to model the disruption in a few CPU-seconds.The best-fitting model quantitatively reproduces the remnant location and the X-shape of the four `plumes'. The normalized residuals of that model to the surface brightness data shows that the peculiar perpendicular (`dog-leg')} stream morphology can only be reproduced if rotation of the dwarf progenitor is included.}
\end{figure}

\section{Future Plans}

The Stellar Tidal Stream Survey has yielded so far an unprecedented sample of bright stellar streams in nearby spiral galaxies, including the discovery of observational analogues to the canonical morphologies found in cosmological simulations of stellar halos. This offers a unique opportunity to study in detail the apparently still dramatic last stages of galaxy assembly in the local Universe and to probe the anticipated estimates of frequency of tidal stellar features from the LCDM paradigm for MW-sized galaxies. 

These discoveries have also enabled first qualitative tests with predictions from N-body models of galaxy disruption/accretion (see Figure 4) based only on the fitting of the sky-projected features available from the imaging. Dynamical analysis of these tidal structures can provide unique views of the dark matter halos (and asymmetries) of their host galaxies. The main degeneracy of modelling streams with imaging data alone is between the orbit and the inclination. Even just a handful of individual kinematic line-of-sight velocities in different parts of the streams can break that degeneracy, especially if the streams has more than a single wrap and if the velocities are on opposite sides of the host galaxy.  In fact, the properties of the host galaxy that we can constrain depend on the morphology of the streams. Assuming some kinematics is available, radial streams (`umbrellas´) are useful to probe the dark matter density profile and slope on very extended radial intervals, from the centre to almost the virial radius of the host galaxy. This is shown in the analysis of the NGC 1097 tidal stream (Amorisco et al. 2015; see Figure 4 upper panel), whose stream is sufficiently unique that one kinematic point alone (the progenitor) was enough. Great circle streams  (like NGC 5907; see Figure~4 top panel) with kinematics can probe the shape of the dark matter halo. Dark matter halos are expected to be triaxial and a large stream sample will allow to test this with a statistical measurement of the shapes of the dark matter halos in the local Universe.

To obtain radial velocities of individual tidal debris stars around nearby galaxies is not yet feasible with the current ground-based facilities. The current state-of-the-art for  spectroscopic follow-up of these features is given by the recent study of the NGC 4449 stream (situated at 4 Mpc) by Toloba et al. (2016). These authors used the Keck-DEIMOS spectrograph to do a first kinematic study of a stellar stream outside the Local Group based on blends of red giant branch stars, including a first spectroscopic metallicity estimate of this stream. However, more robust results using velocities of hundreds of streams stars with better signal-to-noise (S/N ≥20) demands observations with the European Extremely Large Telescope (ELT) and the future MOSAIC instrument (Evans et al. 2018)  early in the next decade. The GLAO\footnote{Ground Layer Adaptive Optics} High Multiplex mode will allow Calcium-Triple line observations of up to 200 objects, providing an exciting new dataset to probe the dark matter halos of  at least 10 galaxies of interest beyond the Local Group.

During the upcoming years, the main objective of the STSS will be to identify those stellar streams  in our cosmic neighborhood that, based on their properties (surface brightness, morphology, orbital orientation, etc), are  the most promising targets to undertake a dynamical study with the MOSAIC instrument. Besides this primary objective, the results of this survey have the potential to tackle a significant number of other topics that are the focus of current astrophysical research (e.g. stellar populations of halos, the resilience of the disks involved in minor mergers, accretion of globular clusters, intra-halo light, induced star formation in streams, near-field cosmology, satellite dynamics, dark matter halo shapes, etc). This research will also provide an essential framework for exploring whether the Milky Way is a template for the archetypal spiral galaxy. In this regard, our survey will be complementary in interpreting the local Galactic archaeological data from the next generation of Galactic surveys (e.g. LSST) and the astrometry mission Gaia in the context of galaxy formation and evolution, providing unique data in order to quantify how typical the Milky Way is with respect to other nearby galaxies of its type.

%
\small  
%
\section*{Acknowledgments}   
%
I thank to the Sociedad Espa\~nola de Astronom\'\i a for the {\it I Premio Javier Gorosabel de Colaboraci\'on ProAm en Astrof\'\i sica} and the astrophotographers of the STSS who have contributed to this project during the last 10 years: R. Jay Gabany, Ken Crawford, Mark Hanson, Karel Teuwen, Johannes Schedler and Adam Block. I also thank Nicola Amorisco, Andrew Cooper, Denis Erkal, Seppo Laine, Giuseppe Donatiello, Emilio G\'alvez and Chris Evans for useful comments. I thank Facundo A. G\'omez for the snapshots from the Auriga cosmological simulations showed in Figure~ 1. I acknowledge support by Sonderforschungsbereich (SFB) 881
``The Milky Way System'' sub-project A2 of the German Research Foundation (DFG) and the Spanish MINECO grant AYA2016-81065-C2-2.
%

%

\begin{thebibliography}{}
\small
%



\bibitem{} {Amorisco, N., Mart{\'{\i}}nez-Delgado, D., Schedler, J. 2015, arXiv:1504.03697}
\bibitem{}{Belokurov, V. et al. 2006, ApJ, 642, L137}
\bibitem{}{Bullock, J. S., Johnston, K. V. 2005, ApJ, 635, 931}
\bibitem{}{Cooper, A. P. et al. 2010, MNRAS, 406, 744}
\bibitem{}{Evans, C. J. et al. 2018 in {\it Early Science with ELTs}, Proceeding IAU Symposium No. 247 (arXiv: 1810.01738)}
\bibitem{}{Foster, C. et al. 2014, MNRAS, 442, 3544}
\bibitem{}{Geha, M. et al. 2017, ApJ, 847, 4}
\bibitem{} {Gibbons, S. L. J., Belokurov, V., Evans, N. W. 2014, MNRAS, 445, 3788}
\bibitem{}{Grand, R. J. J. et al. 2017, MNRAS, 467, 179}
\bibitem{}{Ibata, R., Martin, N.\ F., Irwin, M.\ et al.\ 2007, ApJ, 671, 1591}
\bibitem{}{Johnston, K. V., Sackett, P. D., Bullock, J. S. 2001, ApJ, 557, 137}
\bibitem{}{Johnston, K. V., Bullock, J. S., Sharma, S., et al. 2008, ApJ, 689, 936}
\bibitem{}{ Mart{\'{\i}}nez-Delgado, D., Pe\~narrubia, J., Gabany,  R. J. et al. 2008, ApJ, 689, 184}
\bibitem{}{Mart{\'{\i}}nez-Delgado, D. et al.  2010, ApJ, 140, 962}
\bibitem{}{Mart{\'{\i}}nez-Delgado, D. et al.  2012, ApJ, 748, 24}
\bibitem{}{ Morales, G., Mart{\'{\i}}nez-Delgado, D., Grebel, E. K. et al.  2018, A\&A, 614, 143}
\bibitem{} {Toloba et al. 2016, ApJ, 824, 35}
%
%
\end{thebibliography}
\end{document}